# An integrated language-vision foundation model for conversational diagnostics and triaging in primary eye care

**Running title**: Language-vision foundation model for primary eye care

**Authors**: Zhi Da Soh, PhD [1] †, Yang Bai, PhD [2] †, Kai Yu, PhD [3], Yang Zhou, PhD [2], Xiaofeng Lei, MSc [2], Sahil Thakur, MS [1], Zann Lee, BSc [1], Lee Ching Linette Phang, MSc [1], Qingsheng Peng, MD [1], Can Can Xue, PhD [1], Rachel Shujuan Chong, MD [1, 4], Quan V. Hoang, MD [1, 4, 5], Lavanya Raghavan, MRCS [1], Yih Chung Tham, PhD [1, 6], Charumathi Sabanayagam, MD [1, 4], Wei-Chi Wu, MD, PhD [7, 8] Ming-Chih Ho MD [7, 8], Jiangnan He, PhD [9], Preeti Gupta PhD [1], Ecosse Lamoureux, PhD [1, 10], Seang Mei Saw, MD [1], Vinay Nangia MD [15], Songhomitra Panda-Jonas MD [12, 16], Jie Xu, MD [17], Ya Xing Wang MD [13, 17], Xinxing Xu, PhD [2], Jost B. Jonas, MD [1, 11, 12, 13, 14], Tien Yin Wong, MD[1,13], Rick Siow Mong Goh, PhD [2], Yong Liu, PhD [2] ‡, Ching-Yu Cheng, MD [1, 4, 6, 18] ‡ *

†Contributed equally as first author; ‡ contributed equally as last author; * Corresponding author

**Affiliation**:

1. Singapore Eye Research Institute, Singapore National Eye Centre, Singapore
   Address: 20 College Road, Singapore 169856.

2. Institute of High Performance Computing, Agency of Science, Technology and Research, Singapore
   Address: 1 Fusionopolis Way, Singapore 138632.

3. Department of Radiology, University of Pennsylvania, Philadephia, USA
   Address: Philadephia, PA 19104, USA

4. Ophthalmology & Visual Sciences Academic Clinical Program, Duke-NUS Medical School, Singapore
   Address: 8 College Road, Singapore 169857.

5. Department of Ophthalmology, Columbia University, New York, USA
   Address: 622 W 168th Street, 3rd Floor, New York, NY10032, United States

6. Centre for Innovation and Precision Eye Health, National University of Singapore, Singapore
   Address: 1E Kent Ridge Road, NUHS Tower Block, Level 7, Singapore 119228.

7. Department of Ophthalmology, Chang Gung Memorial Hospital, Linkou Medical Centre, Taoyuan, Taiwan
   Address: No. 123, Dapi Rd., Niaosong District, Kaohsiung City, 83301, Taiwan.

8. College of Medicine, Chang Gung University, Taoyuan, Taiwan
   Address: N0 259, Wenhua 1st road, Guishan District, Taoyuan, 333, Taiwan

9. Shanghai Eye Disease Prevention & Treatment Center, Shanghai Eye Hospital, China
   Address: No. 1440, Hongqiao Road, Changning District, Shanghai, China.

10. Duke-NUS Medical School, National University of Singapore, Singapore.
    Address: 8 College Road, Singapore 169857.

11. Rothschild Foundation Hospital, Institut Français de Myopie, Paris, France
    Address: 25-29 Rue Manin, 75019 Paris, France.

12. Privatpraxis Prof Jonas and Dr. Panda-Jonas
    Address: Adenauerplatz 2, 69115, Heidelberg, Germany

13. Beijing Visual Science and Translational Eye Research Institute, Beijing Tsinghua Changgung Hospital, Tsinghua Medicine, Tsinghua University, Beijing, China
    Address: 30 Shuangqing Road, Haidian District, Beijing, 100190, China.

14. New York eye and Ear Infirmary of Mount Sinai, Icahn School of Medicine at Mount Sinai, New York, United States of America
    Address: 1 Gustave L. Levy Pl, New York, NY 10029, United States.

15. Suraj Eye Institute, Nagpur, India
    Address: Plot No. 559, New Colony, Nagpur, Maharashtra 440001, India.

16. Department of Ophthalmology, Medical Faculty Heidelberg, Heidelberg University, Germany
    Address: Grabengasse 1, 69117 Heidelberg, Germany.

17. Beijing Institute of Ophthalmology, Beijing Tongren Hospital, Capital Meical University, Beijing, China
    Address: Fengtai District, 100054, China.

18. Department of Ophthalmology, National University of Singapore, Singapore
    Address: 1E Kent Ridge Road, NUHS Tower Block, Level 7, Singapore 119228.

**Financial support**:

This study was funded by the National Medical Research Council, Singapore (NMRC/CIRG/1488/2018, NMRC/CIRG33jul-0039 and MOH-CSASI22jul-0001 to CYC) and the Agency for Science, Technology and Research (A*Star) AME Programmatic Funds (grant A20H4b0141 to YL and CYC, and H20H7a0031 CYC and XX). The funding organizations had no role in the design and conduct of this research.

**Declaration of interest**:

All authors declare no conflict-of-interest.

**Corresponding author**:

Professor Ching-Yu Cheng
Email: chingyu.cheng@nus.edu.sg
Address: 1E Kent Ridge Road, NUHS Tower Block, Level 7, Singapore 119228

## Summary

Current deep learning models are mostly task specific and lack a user-friendly interface to operate. We present Meta-EyeFM, a multi-function foundation model that integrates a large language model (LLM) with vision foundation models (VFMs) for ocular disease assessment. Meta-EyeFM leverages a routing mechanism to enable accurate task-specific analysis based on text queries. Using Low Rank Adaptation, we fine-tuned our VFMs to detect ocular and systemic diseases, differentiate ocular disease severity, and identify common ocular signs. The model achieved 100% accuracy in routing fundus images to appropriate VFMs, which achieved ≥82.2% accuracy in disease detection, ≥89% in severity differentiation, ≥76% in sign identification. Meta-EyeFM was 11% to 43% more accurate than Gemini-1.5-flash and ChatGPT-4o LMMs in detecting various eye diseases and comparable to an ophthalmologist. This system offers enhanced usability and diagnostic performance, making it a valuable decision support tool for primary eye care or an online LLM for fundus evaluation.

# Introduction

Vision impairment is the third highest cause of years lived with disability,[1] with significant implications to our mental and socio-economic well-being.[2] Although interventions with good prognosis are available, large proportions of people across different communities worldwide remain undiagnosed for major vision-debilitating eye diseases, such as age-related macular degeneration (AMD) and glaucoma.[3-5] Among other challenges, limited and uneven distribution of skilled eyecare personnel and tertiary-level ophthalmic equipment are significant barriers to disease detection and appropriate referral and treatment.[6]

The advents of artificial intelligence (AI) provide an opportunity to develop automated tools that aid in the screening and detection of eye diseases outside of tertiary care settings using basic equipment in primary care settings. In this regard, several deep learning (DL) algorithms have been approved for diabetic retinopathy (DR) screening,[7,8] and more have demonstrated ophthalmologist-level diagnostic performance.[9,10] However, current DL models were mostly developed to perform a single to few tasks in specific diseases based on fundus photographs obtained from specific cameras. This lacks generalizability and limits its utility in real-world settings where different eye diseases are assessed concurrently, and different brands of fundus camera are used. Thus, a more universal DL models is needed in ophthalmology.

This universal attribute can be obtained from the recently introduced foundation models. These models are first pre-trained on a large corpus of unlabelled data to learn general data features, which enables them to adapt to various downstream tasks without extensive fine-tuning.[11] As compared to traditional DL algorithms, foundation models are more adaptable, scalable, data-efficient, and generalizable.[12,13] Foundation models include vision foundation model (VFM; e.g., RetFound)[14] and large language model (LLM; e.g., GPT-4).[15] In healthcare, VFMs are developed to perform image analysis and captioning, but lack an interactive feature for user's feedback and queries.[14,16] In contrast, LLMs are known for their natural language processing capabilities, which enable them to generate human-like replies to text queries but are less accurate with visual diagnostic tasks.[17] Thus, there remains a need to synergise the complementary strengths of VFMs and LLMs to build a universal DL model that excels in image and text analysis.

Therefore, we aimed to develop a language-vision foundation model, Meta-EyeFM, that combines the capabilities of VFMs and LLMs for ocular health assessment. We developed VFMs to assess the leading causes of vision impairment (e.g., DR, AMD, myopic macular degeneration [MMD], cataract, and glaucoma), covering tasks such as disease detection, severity differentiation, and signs identification. The VFMs further evaluated systemic diseases associated with ocular health, including diabetes, hypertension (HTN), and chronic kidney disease (CKD). The VFMs are integrated within a LLM interface to facilitate conversational AI diagnostics (**Figure 1**).

**Figure 1. Overview of Meta-EyeFM as an ocular and systemic disease classifier and large language model**

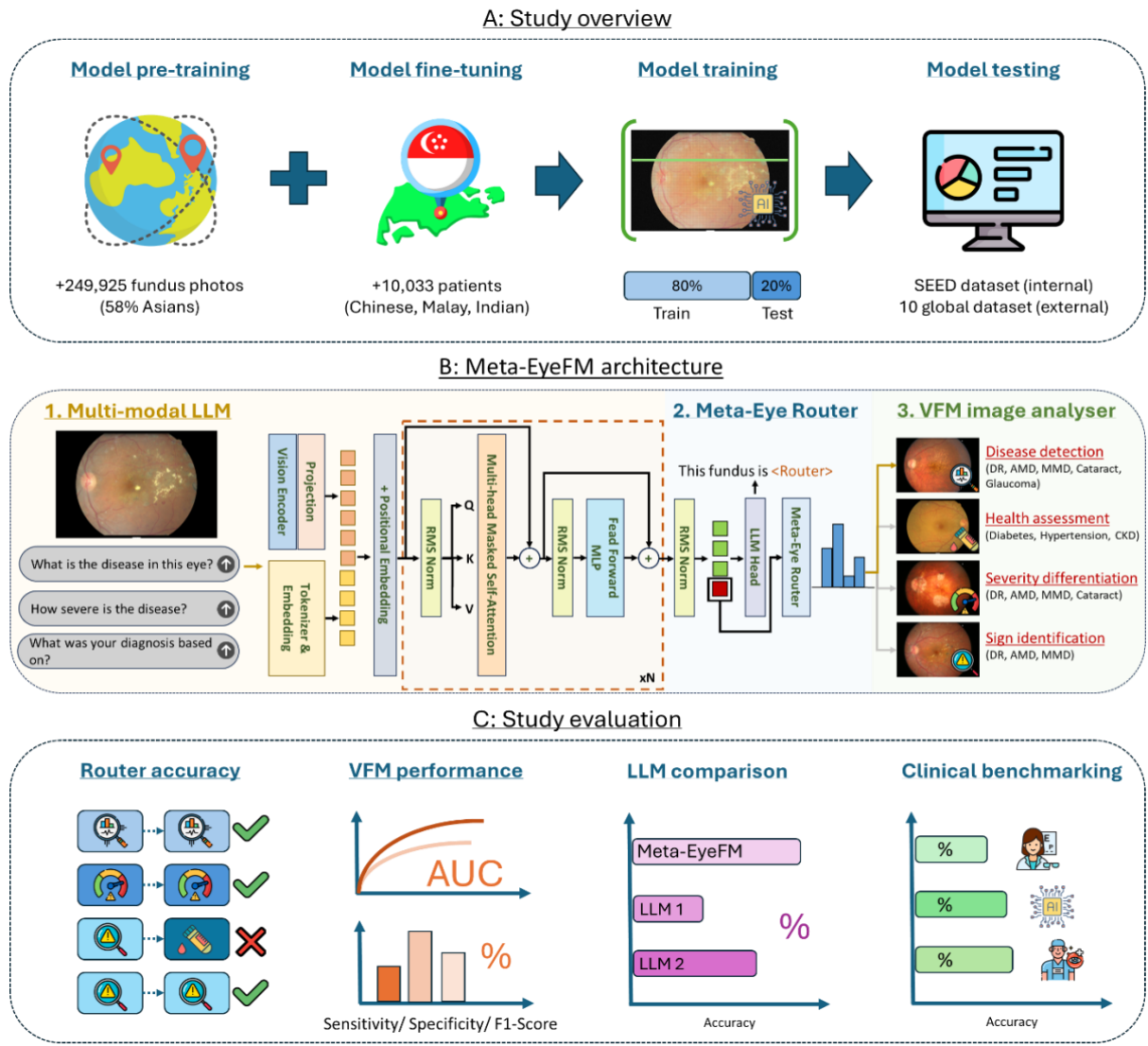

Footnote: When a user wants to identify the disease present in a fundus photograph, our LLM will route the query and image to the VFMs trained to detect ocular and systemic diseases. Likewise, if a user asks about the severity of an ocular disease, the LLM will route the query and image to the VFMs trained to differentiate disease severity (B).

Acronym: VFM, Vision foundation model; LLM, Large language model; DR, Diabetic retinopathy; AMD, Age-related macular degeneration; MMD, Myopic macular degeneration; CKD, Chronic kidney disease; AUC, Area under the receiver operating curve.

## Methods

**Data used in algorithm development**

For model pre-training, we utilized 249,925 fundus photographs (41.3% Caucasians; 58.7% Asians) from 22 population-based studies across 4 continents (**Table S1**). The top contributors included the Netherlands (40.5%) and Singapore (40.7%).

For model fine-tuning, we utilized baseline data (18,736 fundus photographs) from the Singapore Epidemiology of Eye Disease (SEED) Study (2004-2011), a long-term population-based cohort that recruited Malay, Indian, and Chinese adults aged ≥40 years.[18] After written informed consent, study participants underwent a comprehensive ocular examination by ophthalmology-trained investigators and fundus photographs were obtained after pupil dilation.[18] Disease diagnosis were further confirmed by sub-specialty trained research fellows and adjudicated by a panel of senior specialists if required. Details of the SEED data used are summarized in **Table S2**. We further included 65,664 fundus photographs from 10 private and open-source datasets for external testing (**Table S3**).

We excluded glaucoma suspects and standardized disease classification across fine-tuning and external testing datasets by reclassifying DR into non-referable and referable (moderate non-proliferative DR or worse) and MMD into non-referable and referable (category 2 or worse and any plus signs). We obtained ethics approval from the SingHealth Centralized Institutional Board and conducted our study according to the tenets of the Declaration of Helsinki. The funding organizations had no role in the design and conduct of this research.

**Overview of model architecture**

Meta-EyeFM consists of two main components: a multi-modal LLM and eight task-specific VFMs. Our LLM dynamically routes input images to the appropriate VFMs based on user queries rather than generates answers directly (in which LLMs are not optimized for in diagnostic contexts). This design trains our model to process fundus photographs and user queries to understand image patterns and user intentions. This intelligent routing system is essential in managing diverse diagnostic requirements and aims to improve diagnostic accuracy and operational efficiency.

**Development of the multi-modal large language model**

We employed a fine-tuning approach based on Low-Rank Adaptation (LoRA) to develop our multi-modal LLM into an effective router.[19] The LoRA technique enhanced the model's capabilities by adding essential domain knowledge without updating all the original LLM parameters, thereby optimizing efficiency. As the primary foundational LLM, initially based on models like LLaVA,[20] lacked domain knowledge to handle ophthalmology related queries, we implemented a supervised fine-tuning approach to integrate ophthalmology specific knowledge into the LLM training process.

The LoRA technique introduced additional network layers that form a bypass path to the original LLM. During training, the parameters of the LLM remained fixed, with only the matrices $A$ (for reduction) and $B$ (for expansion) undergoing training. The dimensionality-reducing matrix $A$ is initialized with a random Gaussian distribution, while the dimensionality-expanding matrix $B$ is initialized as a zero matrix. This process is formulated as $y = W_0 x + BAx$, where $x$ and $y$ are the input and output, and $W_0$ is the pre-trained weights of the original multi-model LLM.

**Query types and input structure**

Our LLM was trained to handle three types of user queries with reference to an input fundus photograph. The first type of query pertains to disease detection (e.g., "Could you identify the specific disease present in this fundus image?", "Is there any disease in this eye?", and "What is wrong with this eye?"). The second type of query pertains to disease severity (e.g., "How severe is the disease?" and "What is the extent of the condition in this eye?"). The third type of query pertains to disease signs (e.g., "What are the abnormalities seen in this eye?", "How did you make your diagnosis?", and "Are there signs of other diseases in this eye?"). The disease identified from the first type of query is appended to the second and third type of query to provide contextual knowledge.

The SEED dataset was used to fine-tune our multi-modal LLM. For each sample in the dataset, the three types of queries were used in the training set. Since multiple query variants were sampled randomly for each type, the number of training samples was tripled, ensuring the model learns to handle a wide range of diagnostic queries and routing decisions.

**Multi-modal large language model routing**

We introduced the embedding-as-router paradigm to equip the LLM with routing capabilities.[21] This involved expanding the original language model's vocabulary with a novel token, <Router>, which served as a marker for the routing prediction task. Upon receiving a user text query $X_t$ and a corresponding fundus image $X_{img}$, the multi-modal LLM $F$ processed these inputs to produce a text response $\hat{y}$, expressed as: $\hat{y} = F(X_{img}, X_t)$.

When a routing decision is required, the output $\hat{y}$ included the <Router> token. The LLM's final layer embedding $h_{router}$, associated with the <Router> token was extracted and processed through a multi-layer perceptron (MLP) projection layer to yield the routing results $\hat{y}_{router} \in R^D$. Here, $D$ denotes the number of expert models. The $i^{th}$ routing probability $\hat{y}^i_{router}$ was optimized using Binary Cross Entropy (BCE) loss.

**Development of vision foundation model**

Our VFM was pretrained using specific configuration of the masked autoencoder,[22] and the initial model weights were based on the RetFound pre-trained weights to leverage on its extensive prior knowledge in fundus imaging.[14] This autoencoder composed of an encoder and a decoder. The encoder employed a large vision Transformer (ViT-large) architecture, featuring 24 Transformer blocks and an embedding vector size of 1,024. In contrast, the decoder employed a small vision

Transformer (ViT-small) with eight Transformer blocks and an embedding vector size of 512. After pretraining, the model weights were fine-tuned using the SEED dataset. Detailed description is provided in **Annex A**.

**Statistical analysis**

First, we computed the area under the receiver operating curve (AUC), area under the precision recall curve (AUPRC), accuracy, sensitivity, specificity and F1-score to assess the classification performance of Meta-EyeFM. Classification thresholds were selected based on Youden's Index.[23] For multi-class classification, we compared a particular class with all other classes of the disease. Random few-shots analyses were performed to evaluate the classification performance of Meta-EyeFM when fine-tuned with 10%, 30% and 50% of available data.

Second, we compared the performance of Meta-EyeFM with Gemini-1.5-flash (Google) and ChatGPT-4o (Open-AI) in detecting DR, AMD, MMD and glaucoma. For each disease, 100 cases and 100 controls were randomly selected for evaluation. All models were assessed with a base prompt (1) *"What disease is in this image?"*. We further evaluated Gemini-1.5-flash and ChatGPT-4o with more detailed prompts (2) *"Given a label set (DR, AMD, MMD, glaucoma): What is the disease in this image?"*, and prompt (3) that asked specifically for the disease in each dataset (e.g., for CGMH, *"Is MMD presents in this image"*). Accuracy, sensitivity, specificity and the F1-Score were computed based on keywords generated by the LLMs.

Third, we benchmarked the performance of Meta-EyeFM to four ophthalmologists with three to over ten years of clinical experience (2 resident level; 2 consultant level) and four optometrists with three to five years of experience (1 Diploma, 2 Bachelor, 1 Master degree). We randomly selected 60 images (68 diagnostic labels, including normal, DR, AMD, MMD, cataract; glaucoma) from the SEED dataset for evaluation. The graders were unaware of the possible cases and graded the images separately in a single setting. Accuracy, sensitivity, specificity and the F1-Score were evaluated. All statistical analyses were performed using STATA version 17 (STATA Corp, Texas, USA) and all P-values presented are two-sided and statistically significant if <0.05.

## Results

**Ocular disease detection**

Our LLM achieved 100% accuracy in routing fundus photographs to the appropriate VFMs based on inputted queries. In internal testing, Meta-EyeFM achieved an AUC of 97.4% (95% confidence interval [CI] 95.7, 99.0) in detecting referable DR, 91.2% (95% CI 88.8, 93.7) for AMD, 98.8% (95% CI 98.4, 99.2) for referable MMD, 94.2% (95% CI 92.0, 96.3) for glaucoma and 93.9% (95% CI 92.8, 94.9) for visually significant cataract with accuracy ≥82.2% and F1-scores ≤0.814 (**Figure 2; Table S4**). In random few-shots analysis, AUC declined ≤2.9% when 50% of available data were used for fine tuning (**Table S5**).

In external testing, Meta-EyeFM achieved AUCs ≥81% in detecting referable DR, ≥81% for AMD, ≥85% for referrable MMD, ≥84% for glaucoma, and 94% of visually significant cataract cases. The F1-scores were ≥0.672 for referable DR, ≥0.261 for AMD, ≥0.899 for referable MMD, ≥0.390 for glaucoma and 0.279 for visually significant cataract (**Table S6**).

**Figure 2. Diagnostic performance of Meta-EyeFM in ocular and systemic disease detection in internal testing**

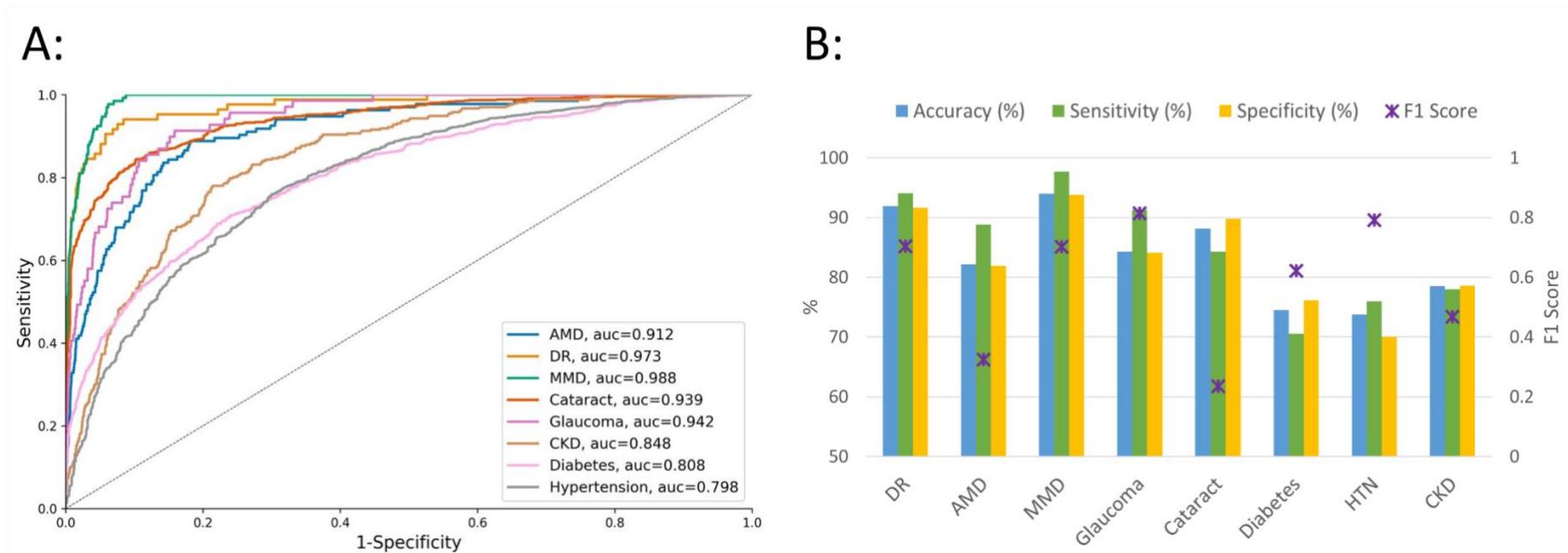

Footnote: Area under the receiver operating curve (A) and diagnostic metric (B) of Meta-EyeFM. Referable DR includes moderate NPDR, severe NPDR and PDR. Referable MMD includes Cat 2 to Cat 4 and all plus sign (including Cat 1+).

Acronym: AUC, Area under the receiver operating curve; AUPRC, Area under the precision recall curve; DR, Diabetic retinopathy; AMD, Age-related macular degeneration; MMD, Myopic macular degeneration; CKD, Chronic kidney disease.

**Ocular disease severity differentiation**

In DR cases, the AUC ranged between 82% (95% CI 76.2, 87.7) for mild non-proliferative DR and 98.9% (95% CI 98.2, 99.7) for proliferative DR with accuracy ≥82.5% and F1-scores ≥0.363(**Figure 3; Table S7**). In AMD cases, the AUC was 91.6% (95% CI 88.6, 93.8) for early AMD and 99.2% for late AMD with F1-scores ≥0.148. In MMD cases, the AUC ranged between 95.4% for fundus tessellation (Cat 1) and 88.7% for patch chorioretinal atrophy and macular atrophy (MMD Cat 3/4) with F1-scores ≥0.379. The AUC was 96% in detecting visually significant cataract with F1-score of 0.387.

**Figure 3. Diagnostic performance of meta-EyeFM in ocular disease severity classification in internal testing.**

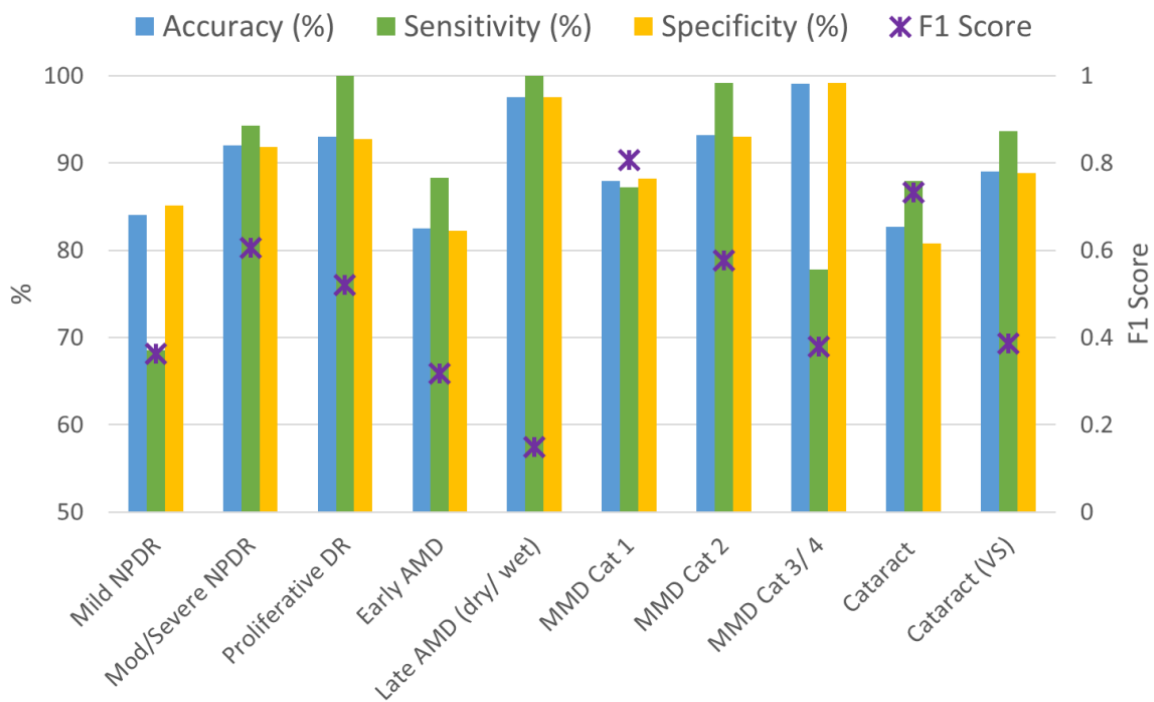

Acronym: AUC, Area under the receiver operating curve; NPDR, non-proliferative diabetic retinopathy; PDR, Proliferative diabetic retinopathy; AMD, Age-related macular degeneration; MMD, Myopic macular degeneration; VS, Visually significant.

## Ocular sign identification

Meta-EyeFM had an AUC of 82% for identifying microaneurysms, 99% for laser scars, 95.6% for hard exudates, 93.4% for soft exudates, and 98.7% for macular oedema (**Figure 4; Table S8**). The accuracy and F1-score were 76.3% and 0.683 for microaneurysms, and ≥81.9% and ≥0.302 for the other signs respectively. In AMD cases, the AUC was 77.9% for drusen with F1-score 0.901, and 89.3% for pigmentary abnormalities with F1-score 0.483. In MMD, the AUC was 95.3% for fundus tessellation with F1-score 0.905, and 97.1% for diffuse retinal atrophy with F1-score 0.679.

**Figure 4. Diagnostic performance of meta-EyeFM in detecting common ocular signs in internal testing**

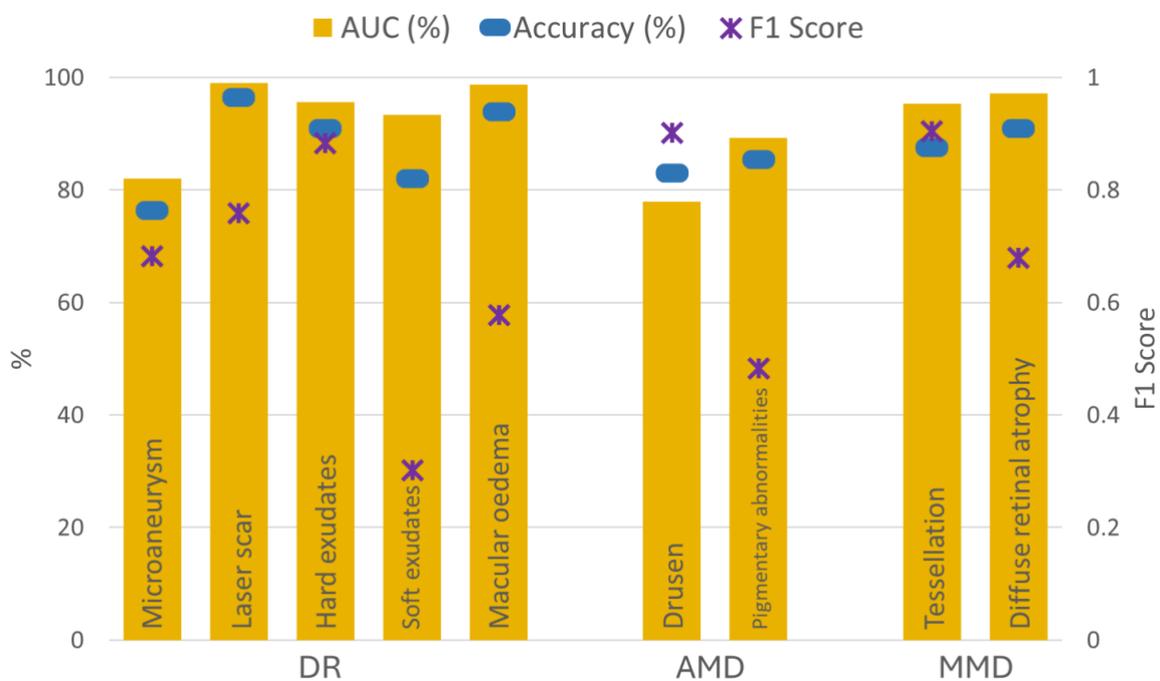

Acronym: AUC, Area under the receiver operating curve; DR, Diabetic retinopathy; AMD, Age-related macular degeneration; MMD, Myopic macular degeneration.

Footnote: Referable DR includes moderate NPDR, severe NPDR and PDR. Referable MMD includes Cat 2 to Cat 4 and all plus sign (including Cat 1+).

**Systemic disease prediction**

In internal testing, Meta-EyeFM achieved AUC of 80.8% (95% CI 79.1, 82.6) in detecting DM, 79.8% (95% CI 78.1, 81.5) for hypertension, and 84.8% (95% CI 82.8, 86.8) for CKD cases with accuracy ≥73.8% and F1-scores ≥0.468 (**Figure 2; Table S4**). In random few-shots analysis, AUC declined ≤1.5% when 50% of data were utilized in fine-tuning (**Table S5**). In external testing, the AUCs were ≥65% for DM, ≥67% for hypertension, and ≥59% for CKD cases with accuracy ≥39.7% and F1-scores ≥0.059 (**Table S6**).

**Comparison with other large language models**

As compared to Gemini-1.5-flash and GPT-4o using the same prompt, the accuracy of Meta-EyeFM was 6% to 21% higher for DR, 25% to 26.5% higher for AMD, 41% to 41.5% higher for MMD, and 30 to 32% higher for glaucoma (**Figure 5; Table S9**). This superior performance by Meta-EyeFM was similarly observed when more detailed prompts were used for Gemini-1.5-flash and GPT-4o (**Figure 5; Table S9**).

**Figure 5. Classification performance of Meta-EyeFM as compared to Gemini and GPT-4o in detecting ocular diseases in external test datasets**

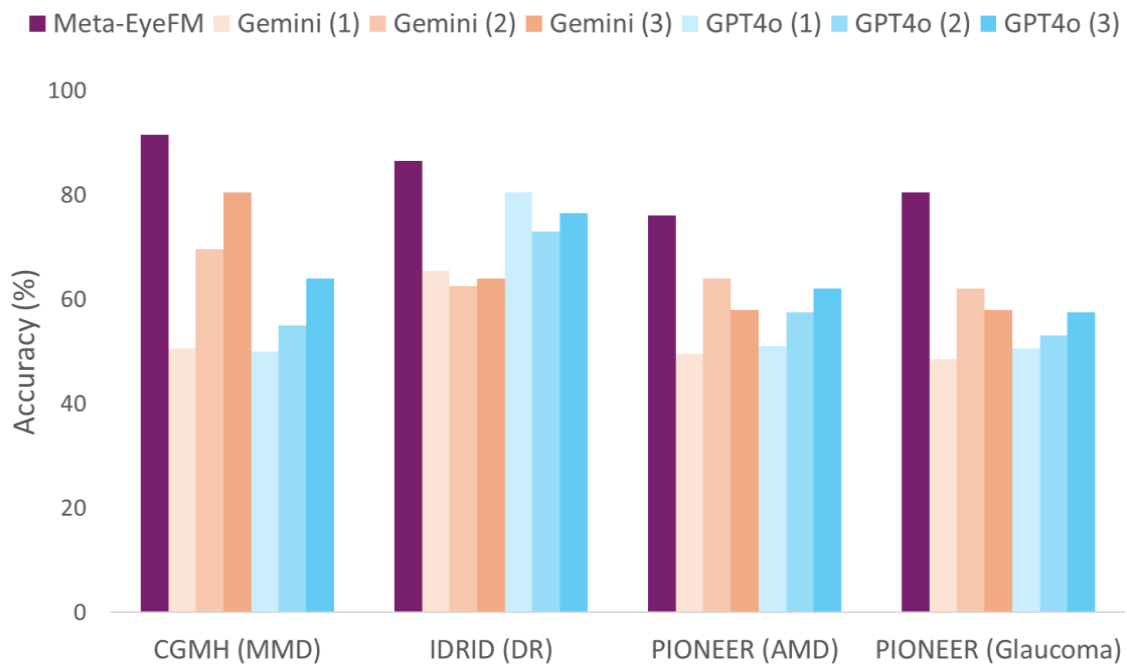

Footnote: (1) denotes prompt "what is the disease in this image"; (2) denotes prompt "given a label set (DR, AMD, MMD, glaucoma): what is the disease in this image"' (3) denotes prompt that asked for the presence of a specific disease (e.g., Is DR or AMD or MMD or glaucoma presents in this image".

Acronym: CGMH, Chang Gung Memorial Hospital; IDRiD, Indian Diabetic Retinopathy Dataset; PIONEER, Population Health and Age-related Sensory Decline Profile; MMD, Myopic macular degeneration; DR, Diabetic retinopathy; AMD, Age-related macular degeneration.

**Clinical benchmarking**

Overall, the classification performance of Meta-EyeFM (F1-score 0.853) was on-par with an ophthalmologist (F1-score 0.857) and higher than the other ophthalmologists (F1-score 0.787 – 0.794) and optometrists (F1-score 0.752 – 0.806; **Table S10**). The difference in F1-Score was largely influenced by glaucoma assessment. Meta-EyeFM achieved a F1-Score of 0.696 in detecting glaucoma, which was slightly lower than an ophthalmologist grader (0.706) but much higher than the other graders (range 0 to 0.333).

## Discussion

In this study, Meta-EyeFM detected major ocular diseases with AUC ≥95% and accuracy ≥82.2%, differentiated ocular disease severity with accuracy ≥89%, identified ocular signs with accuracy ≥76%, and predicted systemic diseases with AUC ≥79.8% and accuracy ≥70.6%. Meta-EyeFM outperformed Gemini-1.5-flash and ChatGPT-4o in detecting DR, AMD, MMD and glaucoma, and was on-par with an ophthalmologist.

Meta-EyeFM is intended for use as a clinical decision support tool to facilitate conversational AI diagnostics using low-cost and readily deployable fundus photographs in primary care settings, unlike other DL models that utilize advanced imaging modalities that are unlikely to be available outside of tertiary care settings.[10,24] Optometrists and even general practitioners are often the initial contacts for mild vision issues. However, referral accuracy varied widely between 48.2% and 88.9% for optometrists and 33% to 67.2% for general practitioners, indicating a need for targeted support.[25] It may also compensate for the scarcity of practitioners in eye screenings and rural areas, and be integrated into telemedicine platforms.

In addition, Meta-EyeFM may operate online like GPT-4o where users can initiate queries on fundus photographs. This aligns with the trend of searching for health-related information online,[26] and utilization of LLMs for eye care purposes is conceivable. A key factor contributing to Meta-EyeFM's superior performance over Gemini-1.5-flash and GPT-4o is its advanced routing capability, which boasts 100% accuracy in directing queries to the most appropriate VFMs. This interactive design distinguish Meta-EyeFM from previous DL algorithms, including RetFound,[14] and mitigates LLM accuracy issues in domain-specific visual diagnostic tasks while enhancing VFM usability.

An integrated decision support tool for primary eye care

We utilized a self-supervised learning approach in developing Meta-EyeFM, which allowed it to maintain its diagnostic accuracy even when 50% of available data were utilized in fine-tuning. This demonstrates data efficiency, which holds potential for including less common eye diseases and ethnic communities without large and well-labelled data. Also, diagnostic performances were generally robust in external testing across diverse datasets, which demonstrate its generalizability and enhances the application prospects of Meta-EyeFM in diverse settings.

The ability to differentiate disease severity provides granularity for disease documentation, monitoring and management. Meta-EyeFM differentiated referrable DR and MMD and further distinguishes mild NPDR and fundus tessellation (MMD Cat 1) among non-referral cases for follow-up. It also differentiated cataract cases with significant vision loss (VA ≤6/18) who will benefit more from medical attention rather than spectacle prescription. This approach right-site referrals and is intended to assist junior optometrists who might be overly cautious in their management.[25]

Previous DL algorithms often overlook disease sign identification, which is important for clinical documentation and disease monitoring.[27] Sign identification in Meta-EyeFM allows user to appraise its diagnostic accuracy and monitor for signs that do not require referral (e.g., small drusen). It further provides insights into AI black-box decision making and complement traditional heatmap visualizations that highlight regions-of-interest but not the specific features within it that influenced its output.[28] This feature is important as clinicians are more willing to adopt AI tools that provide some form of explanation.[29]

We further trained Meta-EyeFM to predict systemic diseases associated with ocular health to provide a more holistic eye assessment. For example, 34.6% of people with DM are affected by DR to some degree but many are unaware due to its insidious nature.[30] Predicting DM from fundus photographs provides a non-invasive method to identify and refer patients with undetected DM for treatment. However, diagnostic improvements are required before utility.

Strengths and limitations.

We utilized a large and diverse population cohort to fine-tune Meta-EyeFM to develop a more inclusive AI tool for community use. This cohort reflects the pattern of age-related eye diseases among major Asian ethnicities (Chinese, Malay, Indian), which is important given Asia's substantial contribution to global visual impairment rates due to its aging population and healthcare gaps.[31]

However, there are few limitations. First, Meta-EyeFM currently only evaluates major eye diseases and certain categories of severity and signs. Future studies should include less-common diseases. Second, although we added numerous Asian data in addition to Caucasian data in pre-training, Meta-EyeFM still lacks representation (e.g., African, children and adolescents) and collaborative efforts are needed to address this issue. Third, further improvement in diagnostic performances is required before our model can be utilized for screening purposes.

## Conclusion

In summary, Meta-EyeFM is an integrated language-vision foundation model that provides a user-friendly interface for primary eye care, using text and fundus photos to detect a range of major age-related ocular diseases, differentiates ocular diseases severity, identify common ocular signs, and predict systemic diseases associated with ocular health. Meta-EyeFM may act as a clinical decision support tool in primary care settings and as an online eye-specific LLM platform for fundus evaluation.

**References:**

1. Collaborators G. Global, regional, and national incidence, prevalence, and years lived with disability for 354 diseases and injuries for 195 countries and territories, 1990-2017: a systematic analysis for the Global Burden of Disease Study 2017. 2018.
2. Burton MJ, Ramke J, Marques AP, et al. The Lancet global health Commission on global eye health: vision beyond 2020. *The Lancet Global Health.* 2021;9(4):e489-e551.
3. Neely DC, Bray KJ, Huisingh CE, Clark ME, McGwin G, Owsley C. Prevalence of undiagnosed age-related macular degeneration in primary eye care. *JAMA ophthalmology.* 2017;135(6):570-575.
4. Soh ZD, Yu M, Betzler BK, et al. The global extent of undetected glaucoma in adults: a systematic review and meta-analysis. *Ophthalmology.* 2021;128(10):1393-1404.
5. Chua J, Lim B, Fenwick EK, et al. Prevalence, Risk Factors, and Impact of Undiagnosed Visually Significant Cataract: The Singapore Epidemiology of Eye Diseases Study. *PLOS ONE.* 2017;12(1):e0170804.
6. Organization WH. World report on vision. 2019.
7. Gulshan V, Peng L, Coram M, et al. Development and validation of a deep learning algorithm for detection of diabetic retinopathy in retinal fundus photographs. *jama.* 2016;316(22):2402-2410.
8. Ting DSW, Cheung CY-L, Lim G, et al. Development and validation of a deep learning system for diabetic retinopathy and related eye diseases using retinal images from multiethnic populations with diabetes. *Jama.* 2017;318(22):2211-2223.
9. Grassmann F, Mengelkamp J, Brandl C, et al. A deep learning algorithm for prediction of age-related eye disease study severity scale for age-related macular degeneration from color fundus photography. *Ophthalmology.* 2018;125(9):1410-1420.
10. Lee CS, Baughman DM, Lee AY. Deep learning is effective for classifying normal versus age-related macular degeneration OCT images. *Ophthalmology Retina.* 2017;1(4):322-327.
11. Krishnan R, Rajpurkar P, Topol EJ. Self-supervised learning in medicine and healthcare. *Nature Biomedical Engineering.* 2022;6(12):1346-1352.
12. Chen T, Kornblith S, Norouzi M, Hinton G. A simple framework for contrastive learning of visual representations. Paper presented at: International conference on machine learning2020.
13. Chen T, Kornblith S, Swersky K, Norouzi M, Hinton GE. Big self-supervised models are strong semi-supervised learners. *Advances in neural information processing systems.* 2020;33:22243-22255.
14. Zhou Y, Chia MA, Wagner SK, et al. A foundation model for generalizable disease detection from retinal images. *Nature.* 2023;622(7981):156-163.
15. Nori H, King N, McKinney SM, Carignan D, Horvitz E. Capabilities of gpt-4 on medical challenge problems. *arXiv preprint arXiv:230313375.* 2023.
16. Christensen M, Vukadinovic M, Yuan N, Ouyang D. Vision–language foundation model for echocardiogram interpretation. *Nature Medicine.* 2024:1-8.
17. Wu C, Lei J, Zheng Q, et al. Can gpt-4v (ision) serve medical applications? case studies on gpt-4v for multimodal medical diagnosis. *arXiv preprint arXiv:231009909.* 2023.
18. Majithia S, Tham Y-C, Chee M-L, et al. Cohort profile: the Singapore epidemiology of eye diseases study (seed). *International journal of epidemiology.* 2021;50(1):41-52.
19. Hu EJ, Shen Y, Wallis P, et al. Lora: Low-rank adaptation of large language models. *arXiv preprint arXiv:210609685.* 2021.
20. Liu H, Li C, Wu Q, Lee YJ. Visual instruction tuning. *Advances in neural information processing systems.* 2024;36.

21. Lai X, Tian Z, Chen Y, et al. Lisa: Reasoning segmentation via large language model. Paper presented at: Proceedings of the IEEE/CVF Conference on Computer Vision and Pattern Recognition2024.
22. He K, Chen X, Xie S, Li Y, Dollár P, Girshick R. Masked autoencoders are scalable vision learners. Paper presented at: Proceedings of the IEEE/CVF conference on computer vision and pattern recognition2022.
23. WJ Y. Index for rating diagnostic tests. *Cancer.* 1950;3:32-35.
24. Wang D, Wang L. On OCT image classification via deep learning. *IEEE Photonics Journal.* 2019;11(5):1-14.
25. Carmichael J, Abdi S, Balaskas K, Costanza E, Blandford A. Assessment of optometrists' referral accuracy and contributing factors: A review. *Ophthalmic and Physiological Optics.* 2023;43(5):1255-1277.
26. Kuehn BM. More than one-third of US individuals use the Internet to self-diagnose. *Jama.* 2013;309(8):756-757.
27. Optometrists. Co. College of Optometrists guideline C143: Communication, Partnership and Teamwork. https://www.college-optometrists.org/clinical-guidance/guidance/communication,-partnership-and-teamwork/working-with-colleagues. Accessed 29 May, 2024.
28. Ghassemi M, Oakden-Rayner L, Beam AL. The false hope of current approaches to explainable artificial intelligence in health care. *The Lancet Digital Health.* 2021;3(11):e745-e750.
29. Panigutti C, Beretta A, Giannotti F, Pedreschi D. Understanding the impact of explanations on advice-taking: a user study for AI-based clinical Decision Support Systems. Paper presented at: Proceedings of the 2022 CHI Conference on Human Factors in Computing Systems2022.
30. Yau JW, Rogers SL, Kawasaki R, et al. Global prevalence and major risk factors of diabetic retinopathy. *Diabetes care.* 2012;35(3):556-564.
31. Bourne RR, Flaxman SR, Braithwaite T, et al. Magnitude, temporal trends, and projections of the global prevalence of blindness and distance and near vision impairment: a systematic review and meta-analysis. *The Lancet Global Health.* 2017;5(9):e888-e897.

**Supplemental information**

**Acknowledgement**

This study was funded by the National Medical Research Council, Singapore (NMRC/CIRG/1488/2018, NMRC/CIRG33jul-0039 and MOH-CSASI22jul-0001 to CYC) and the Agency for Science, Technology and Research (A*Star) AME Programmatic Funds (grant A20H4b0141 to YL and CYC, and H20H7a0031 CYC and XX). The funding organizations had no role in the design and conduct of this research.

The authors would like to acknowledge Sarah Shwu Huey Tan and Zuriati Binte Abdullah for their help in data pre-processing.

**Author contribution**

**Zhi Da Soh**: Conceptualization, data curation, data annotation, statistical analysis, data visualization, writing – original draft; **Yang Bai**: Conceptualization, algorithm development and testing, statistical analysis, writing – original draft; **Kai Yu**: Algorithm development and testing, writing – review & editing; **Yang Zhou**: Conceptualization, writing – review & editing; **Xiaofeng Lei**: Data preparation, writing – review & editing; **Sahil Thakur**: Data annotation, writing – review & editing; **Zann Lee**: Data curation, data annotation, writing – review & editing; **Lee Ching Linette Phang**: Data annotation, writing – review & editing; **Qingsheng Peng**: Data annotation, writing – review & editing; **Can Can Xue**: Data annotation, writing – review & editing; **Rachel Shujuan Chong**: Data annotation, writing – review & editing; **Quan V. Hoang**: Data annotation, writing – review & editing; **Lavanya Raghavan**: Data annotation, writing – review & editing; **Yih Chung Tham**: writing – review & editing; **Wei-Chi Wu**: Data provision, writing – review & editing; **Ming-Chih Ho**: Data preparation, writing – review & editing; **Jiangnan He**: Data provision, writing – review & editing; **Preeti Gupta** – Data preparation, writing – review & editing; **Ecosse Lamoureux** – Data provision, writing – review & editing; **Seang Mei Saw**: Data provision, writing – review & editing; **Vinay Nangia**: Data provision, writing – review & editing; **Songhomitra Panda-Jonas**: Data provision, writing – review & editing; **Jie Xu**: Data preparation, writing – review & editing; **Ya Xing Wang**: Data provision, writing – review & editing;

**Xinxing Xu**: writing – review & editing; **Jost B. Jonas**: Data provision, writing – review & editing; **Tien Yin Wong**: writing – review & editing; **Yong Liu**: Conceptualization, supervision, funding acquisition, writing – review & editing; **Ching-Yu Cheng**: Conceptualization, supervision, funding acquisition, writing – review & editing.

**Data sharing statement**

The private fundus photographs and clinical data of participants included in this study are not publicly available due to patient privacy and the data are meant for research purposes only. On reasonable request, de-identified data used in this study may be made available for academic purpose by the Singapore Eye Research Institute (SERI), subjected to approval by the local institutional review board. Data request can be sent to the Data Access Committee at SERI via seri@seri.com.sg. Any data that can be shared will be released via a Research Collaboration Agreement (RCA) for non-commercial research purpose.

**Table S1. The datasets used in Meta-EyeFM pre-training**

| Dataset | Data source | Country | N (Images) | Fundus camera |
|---|---|---|---|---|
| OPS | Private | Singapore | 6284 | TRC-NW8 (Topcon) |
| DPS | Private | Singapore | 4342 | TRC-NW8 (Topcon) |
| DMP | Private | Singapore | 1230 | CR-1 (Canon) |
| SCORM | Private | Singapore | 11845 | CR-1 (Canon) |
| SEED2 | Private | Singapore | 68893 | CR-1 (Canon) |
| SP2 | Private | Singapore | 9139 | CR-1 (Canon) |
| BES | Private | China | 6616 | CR6-45NM (Canon) |
| CIEMS | Private | India | 20404 | FF450 (Carl Zeiss) |
| ODIR-5K | Open source | China | 6377 | Canon, Carl Zeiss, Kowa |
| FIVES | Open source | China | 750 | TRC-NW8 (Topcon) |
| LAG | Open source | China | 4595 | Canon, Carl Zeiss, Topcon |
| GRAPE | Open source | China | 631 | TRC-NW8 (Topcon), CR-2 AF (Canon) CR-2 Plus AF (Canon) |
| REFUGE | Open source | China | 1200 | CR-2 (Canon) Visucam 500 (Carl Zeiss) |
| SUSTech-SYSU | Open source | China | 1219 | TRC-50DX (Topcon) |
| DeepDRiD | Open source | China | 1784 | Not specified |
| CHAKSU | Open source | India | 1345 | Remido, Forus 3Nethra, Bosch |
| Drishti-GS1 | Open source | India | 101 | Not specified |
| STARE | Open source | USA | 397 | TRV-50 (Topcon) |
| HRF | Open source | Germany | 44 | CR-1 (Canon) |
| ACRIMA | Open source | Spain | 705 | TRC (Topcon) |
| AIROGS | Open source | Netherland | 101,267 | Canan, Topcon, Centervue, Optovue |
| Paraguay | Open source | Paraguay | 757 | Not specified |

Acronym: OPS, Outrum Polyclinic Study; DPS, Deep-Phenotyping Study; DMP, Diabetes Management Project; SCORM, Singapore Cohort Study of the Risk Factors of Myopia; SEED2, Singapore Epidemiology of Eye Diseases (6-year follow up visit); SP2, Singapore Prospective Study Program; BES, Beijing Eye Study; CIEMS, Central India Eye & Medical Study; ODIR-5K, Ocular Disease Intelligent Recognition; FIVES, Fundus Image Vessel Segmentation; LAG, Large-scale Attention-based Glaucoma; GRAPE, Glaucoma Real-world Appraisal Progression Ensemble; REFUGE, Retinal Fundus Glaucoma Challenge; SUSTech-SYSU, *not an acronym*; DeepDRiD, Deep Diabetic Retinopathy Image Dataset; Chaksu, *not an acronym*; Drishti-GS1, *not an acronym*; STARE, Structured Analysis of the Retina Dataset; HRF, High Resolution Fundus Image Database; ACRIMA, *not an acronym*; AIROGS, Artificial Intelligence for Robust Glaucoma Screening Challenge.

**Table S2. The SEED1 dataset used in Meta-EyeFM fine-tuning**

| Disease | Classes | N (Images) | | | |
| --- | --- | --- | --- | --- | --- |
| | | Train | Validation | Test | Total |
| Cataract | No | 8881 | 1176 | 1741 | 11798 |
| | Yes | 3446 | 449 | 679 | 4574 |
| | Visually significant | 412 | 66 | 93 | 571 |
| Glaucoma | No | 12883 | 1689 | 2525 | 17097 |
| | Yes | 319 | 55 | 69 | 443 |
| Diabetic retinopathy | No | 3411 | 451 | 672 | 4534 |
| | Mild NPDR | 277 | 39 | 54 | 370 |
| | Moderate NPDR | 235 | 24 | 51 | 310 |
| | Severe NPDR | 20 | 3 | 2 | 25 |
| | PDR | 156 | 26 | 31 | 213 |
| Age-related macular degeneration | No | 13502 | 1782 | 2651 | 17935 |
| | Early AMD | 537 | 74 | 128 | 739 |
| | Late AMD (Dry) | 8 | 2 | 2 | 12 |
| | Late AMD (Wet) | 42 | 4 | 4 | 50 |
| Myopic macular degeneration | No | 8993 | 1191 | 1764 | 11948 |
| | Cat 1/ 1+ | 3817 | 507 | 766 | 5090 |
| | Cat 2/ 2+ | 626 | 87 | 123 | 836 |
| | Cat 3/ 3+ | 31 | 4 | 6 | 41 |
| | Cat 4/ 4+ | 9 | 2 | 3 | 14 |
| Ocular disease sign | Microaneurysm | 1223 | 151 | 263 | 1637 |
| | Laser scars | 213 | 36 | 46 | 295 |
| | Hard exudates | 349 | 44 | 77 | 470 |
| | Macular oedema | 164 | 16 | 34 | 214 |
| | Drusen | 9836 | 1267 | 1907 | 13010 |
| | Pigmentary abnormalities | 809 | 98 | 192 | 1099 |
| | Tessellation | 4529 | 604 | 901 | 6034 |
| | Diffuse retinal atrophy | 671 | 92 | 132 | 895 |
| Systemic diseases | Diabetes | 4237 | 565 | 848 | 5650 |
| | Hypertension | 9107 | 1216 | 1823 | 12146 |
| | Chronic kidney disease | 1653 | 222 | 332 | 2207 |

Footnote: The Modified Airlie House Classification Scheme was used for DR grading, the International Clinical Diabetic Retinopathy Severity Scale for DR classification, the Wisconsin Age-related Maculopathy grading system for AMD grading, the Meta-analysis for Pathological Myopia (META-PM) classification for MMD, and the modified Wisconsin cataract grading system for cataract grading. Visually significant cataract was defined in eyes with late-stage cataracts (i.e., cortical cataract ≥25% or posterior sub-capsular cataract >5% or nuclear sclerosis ≥grade 4) with best-corrected visual acuity worse than Snellen 6/18. Glaucoma was diagnosed based on structural abnormalities of the optic nerve head with corresponding visual field defects. Where reliable results of structural and functional examinations were unavailable, we followed the principles of the International Society of Geographic and Epidemiological Ophthalmology in diagnosing cases. Diabetes was defined as non-fasting plasma glucose ≥11.1 or HbA1C ≥6.5% or patient reported use of diabetes medication. Hypertension was defined as systolic blood pressure ≥150mmHg or diastolic blood pressure ≥90mmHg or patient reported use of hypertension medication. Chronic kidney disease was defined based on estimated glomerular filtration rate less than 60 mL/min/1.73m2 from serum creatinine.

Acronym: NPDR, Non-proliferative diabetic retinopathy; PDR, Proliferative diabetic retinopathy.

Table S3. Datasets used in Meta-EyeFM external testing

| Dataset | Country | Disease | Classes | | N (Images) | Camera |
|---|---|---|---|---|---|---|
| MESSIDOR2 | France | DR | No | | 1017 | TRC-NW6 (Topcon) |
| | | | Mild DPDR | | 270 | |
| | | | Moderate NPDR | | 347 | |
| | | | Severe NPDR | | 75 | |
| | | | PDR | | 35 | |
| IDRiD | India | DR | No | | 168 | VX-10 (Kowa) |
| | | | Mild DPDR | | 25 | |
| | | | Moderate NPDR | | 168 | |
| | | | Severe NPDR | | 93 | |
| | | | PDR | | 62 | |
| APTOS | India | DR | No | | 1805 | Not specified |
| | | | Mild DPDR | | 370 | |
| | | | Moderate NPDR | | 999 | |
| | | | Severe NPDR | | 193 | |
| | | | PDR | | 295 | |
| Glaucoma Fundus | South Korea | Glaucoma | No | | 788 | AFC-330 (Nidek) |
| | | | Yes | | 756 | |
| PIONEER | Singapore | Glaucoma | No | | 1306 | Canon CR-DGi (Canon) |
| | | | Yes | | 190 | |
| | | AMD | No | | 3731 | |
| | | | Early AMD | | 174 | |
| | | | Late AMD (Dry) | | 35 | |
| | | | Late AMD (Wet) | | 21 | |
| CGMH | Taiwan | MMD | Cat 1/ 1+ | | 126 | S8 (Kowa) |
| | | | Cat 2/ 2+ | | 246 | |
| | | | Cat 3/ 3+ | | 112 | |
| | | | Cat 4/ 4+ | | 199 | |
| SESA | China | MMD | Cat 1/ 1+ | | 300 | OCT-1 Atlantis (Topcon) |
| | | | Cat 2/ 2+ | | 80 | |
| | | | Cat 3/ 3+ | | 62 | |
| | | | Cat 4/ 4+ | | 15 | |
| ODIR-5K | China | Normal | yes | | 3751 | Canon, Carl Zeiss, Kowa |
| | | AMD | Yes | | 264 | |
| | | Glaucoma | Yes | | 276 | |
| | | Cataract | Yes | | 200 | |
| | | Diabetes | Yes | | 1597 | |
| | | Hypertension | Yes | | 125 | |
| SP2 | Singapore | Diabetes | No | | 4236 | CR-1 (Canon) |
| | | | Yes | | 3557 | |
| | | Hypertension | No | | 4610 | |
| | | | Yes | | 3081 | |
| UKBB | England | Diabetes | No | | 10773 | 3D OCT 1000 Mark 2 (Topcon) |
| | | | Yes | | 597 | |
| | | Hypertension | No | | 4981 | |
| | | | Yes | | 6256 | |
| | | CKD | No | | 6934 | |
| | | | Yes | | 363 | |

Acronym: MESSIDOR, Methods to Evaluate Segmentation and Indexing Techniques in the field of Retinal Ophthalmology; IDRiD, Indian Diabetic Retinopathy Dataset; APTOS, Asia Pacific Tele-Ophthalmology Society; PIONEER, Population Health and Age-related Sensory Decline Profile; CGMH, Chang Gung Memorial Hospital; SESA, Shanghai Eye Study for Adults; SP2, Singapore Prospective Study Program; ODIR-5K, Ocular Disease Intelligent Recognition; UKBB, United Kingdom Bio-bank.

**Table S4. Diagnostic performance of Meta-EyeFM in ocular and systemic disease detection in internal testing**

|  | AUC, % (95% CI) | AUPRC (%) | Accuracy (%) | Sensitivity (%) | Specificity (%) | F1 score |
|---|---|---|---|---|---|---|
| Referrable DR | 97.4 (95.7, 99.0) | 88.5 | 91.9 | 94.1 | 91.6 | 0.705 |
| AMD | 91.2 (88.8, 93.7) | 46.6 | 82.2 | 88.8 | 81.9 | 0.324 |
| Referrable MMD | 98.8 (98.4, 99.2) | 81.5 | 94.0 | 97.7 | 93.8 | 0.703 |
| Cataract | 93.9 (92.8, 94.9) | 90.2 | 88.1 | 84.3 | 89.8 | 0.814 |
| Glaucoma | 94.2 (92.0, 96.3) | 46.4 | 84.3 | 91.3 | 84.1 | 0.236 |
| Diabetes | 80.8 (79.1, 82.6) | 69.2 | 74.5 | 70.6 | 76.2 | 0.622 |
| Hypertension | 79.8 (78.1, 81.5) | 85.8 | 73.8 | 76.0 | 70.0 | 0.792 |
| CKD | 84.8 (82.8, 86.8) | 44.7 | 78.5 | 78.0 | 78.6 | 0.468 |

Footnote: Referable DR includes moderate NPDR, severe NPDR and PDR. Referable MMD includes Cat 2 to Cat 4 and all plus sign (including Cat 1+).

Acronym: AUC, Area under the receiver operating curve; CI, Confidence interval; AUPRC, Area under the precision recall curve; DR, Diabetic retinopathy; AMD, Age-related macular degeneration; MMD, Myopic macular degeneration; CKD, Chronic kidney disease.

**Table S5. Random few-shots analysis in detecting ocular and systemic diseases with Meta-EyeFM in internal testing**

|  | 10% of total fine-tune data | | | | 30% of total fine-tune data | | | | 50% of total fine-tune data | | | |
| --- | --- | --- | --- | --- | --- | --- | --- | --- | --- | --- | --- | --- |
|  | AUC (%) | AUPRC (%) | Accuracy (%) | F1 score | AUC (%) | AUPRC (%) | Accuracy (%) | F1 score | AUC (%) | AUPRC (%) | Accuracy (%) | F1 score |
| Referrable DR | 94.5 | 79.6 | 91.1 | 0.695 | 95.9 | 82.7 | 92.8 | 0.758 | 95.4 | 82.0 | 92.2 | 0.749 |
| AMD | 78.0 | 17.5 | 61.2 | 0.029 | 89.0 | 40.8 | 83.2 | 0.253 | 89.4 | 42.5 | 82.1 | 0.304 |
| Referrable MMD | 95.9 | 45.4 | 87.2 | 0.263 | 96.2 | 46.7 | 89.8 | 0.273 | 95.9 | 41.8 | 89.6 | 0.275 |
| Cataract | 92.6 | 88.0 | 85.6 | 0.767 | 93.4 | 89.5 | 87.2 | 0.799 | 93.6 | 89.8 | 88.5 | 0.801 |
| Glaucoma | 82.9 | 19.4 | 62.6 | 0.107 | 90.3 | 42.6 | 82.1 | 0.326 | 93.6 | 46.1 | 90.8 | 0.396 |
| Diabetes | 76.3 | 61.5 | 70.1 | 0.439 | 78.8 | 65.8 | 72.9 | 0.540 | 79.4 | 67.4 | 74.9 | 0.539 |
| Hypertension | 76.5 | 83.1 | 69.3 | 0.790 | 77.5 | 83.6 | 71.2 | 0.798 | 78.3 | 84.6 | 73.4 | 0.801 |
| CKD | 82.3 | 38.6 | 67.3 | 0.326 | 83.3 | 40.0 | 74.5 | 0.249 | 83.7 | 41.6 | 75.5 | 0.326 |

Footnote: Referable DR includes moderate NPDR, severe NPDR and PDR. Referable MMD includes Cat 2 to Cat 4 and all plus sign (including Cat 1+).

Acronym: AUC, Area under the receiver operating curve; AUPRC, Area under the precision recall curve; DR, Diabetic retinopathy; AMD, Age-related macular degeneration; MMD, Myopic macular degeneration; CKD, Chronic kidney disease.

**Table S6. Diagnostic performance of Meta-EyeFM in ocular disease detection in external testing**

|  | AUC % (95% CI) | AUPRC (%) | Accuracy (%) | Sensitivity (%) | Specificity (%) | F1 score |
|---|---|---|---|---|---|---|
| **Referrable DR** | | | | | | |
| • APTOS | 99.0 (98.4, 99.6) | 99.2 | 95.9 | 93.2 | 98.7 | 0.959 |
| • IDRID | 91.9 (86.7, 97.2) | 96.1 | 87.4 | 79.7 | 100.0 | 0.887 |
| • MESSIDOR2 | 80.7 (76.0, 85.4) | 74.2 | 82.7 | 66.9 | 88.4 | 0.672 |
| **AMD** | | | | | | |
| • PIONEER | 80.5 (77.1, 84.0) | 24.5 | 75.9 | 73.5 | 76.0 | 0.261 |
| • ODIR-5K | 90.8 (88.4, 93.2) | 56.6 | 86.2 | 82.6 | 86.4 | 0.337 |
| **Referrable MMD** | | | | | | |
| • CGMH | 85.4 (82.7, 88.0) | 91.1 | 85.4 | 90.8 | 71.1 | 0.899 |
| • SESA | 100 (100, 100) | 100 | 100.0 | 100.0 | 100.0 | 1.000 |
| **Glaucoma** | | | | | | |
| • PIONEER | 84.0 (80.4, 87.6) | 49.3 | 82.0 | 67.9 | 84.0 | 0.489 |
| • GF | 90.2 (87.3, 93.1) | 89.9 | 81.7 | 79.8 | 83.5 | 0.811 |
| • ODIR-5K | 93.8 (91.8, 95.7) | 48.0 | 87.5 | 89.9 | 87.4 | 0.390 |
| **Cataract** | | | | | | |
| • ODIR-5K | 94.4 (92.2, 96.6) | 36.8 | 84.6 | 92.5 | 84.4 | 0.279 |
| **Diabetes** | | | | | | |
| • ODIR-5K | 68.5 (66.9, 70.1) | 48.9 | 71.5 | 48.0 | 79.7 | 0.464 |
| • SP2 | 69.1 (67.9, 70.3) | 65.8 | 64.2 | 56.9 | 70.3 | 0.591 |
| • UKBB | 64.5 (62.0, 66.9) | 11.3 | 65.5 | 54.6 | 66.1 | 0.142 |
| **Hypertension** | | | | | | |
| • ODIR-5K | 68.7 (63.5, 73.9) | 3.9 | 39.7 | 95.2 | 38.6 | 0.059 |
| • SP2 | 80.6 (79.6, 81.7) | 71.8 | 73.5 | 75.4 | 72.2 | 0.695 |
| • UKBB | 66.5 (65.5, 67.5) | 68.2 | 63.0 | 66.6 | 58.4 | 0.667 |
| **CKD** | | | | | | |
| • UKBB | 58.7 (55.6, 61.9) | 6.7 | 41.8 | 74.9 | 40.1 | 0.113 |

Footnote: Referable DR includes moderate NPDR, severe NPDR and PDR. Referable MMD includes Cat 2 to Cat 4 and all plus sign (including Cat 1+).

Acronym: AUC, Area under the receiver operating curve; AUPRC, Area under the precision recall curve; DR, Diabetic retinopathy; AMD, Age-related macular degeneration; MMD, Myopic macular degeneration; CKD, Chronic kidney disease.

**Table S7. Diagnostic performance of meta-EyeFM in ocular disease severity classification**

|  | AUC, % (95% CI) | AUPRC (%) | Accuracy (%) | Sensitivity (%) | Specificity (%) | F1 score |
|---|---|---|---|---|---|---|
| **DR** |  |  |  |  |  |  |
| • Mild NPDR | 82.0 (76.2, 87.7) | 27.3 | 84.0 | 68.5 | 85.1 | 0.363 |
| • Mod/Severe NPDR | 96.4 (94.2, 98.7) | 73.7 | 92.0 | 94.3 | 91.8 | 0.606 |
| • PDR | 98.9 (98.2, 99.7) | 84.8 | 93.0 | 100.0 | 92.7 | 0.521 |
| **AMD** |  |  |  |  |  |  |
| • Early | 91.6 (88.6, 93.8) | 49.0 | 82.5 | 88.3 | 82.2 | 0.317 |
| • Late (dry/ wet) | 99.2 (98.5, 99.9) | 32.5 | 97.5 | 100.0 | 97.5 | 0.148 |
| **MMD** |  |  |  |  |  |  |
| • Cat 1 | 95.4 (94.5, 96.1) | 90.4 | 87.9 | 87.2 | 88.2 | 0.806 |
| • Cat 2 | 98.6 (98.1, 99.0) | 74.7 | 93.2 | 99.2 | 93.0 | 0.576 |
| • Cat 3 to 4 | 88.7 (65.8, 100) | 48.4 | 99.1 | 77.8 | 99.2 | 0.379 |
| **Cataract** |  |  |  |  |  |  |
| • Yes | 91.7 (90.6, 93.0) | 82.4 | 82.7 | 87.9 | 80.8 | 0.733 |
| • Visually significant | 96.0 (94.2, 98.3) | 60.1 | 89.0 | 93.6 | 88.8 | 0.387 |

Acronym: AUC, Area under the receiver operating curve; AUPRC, Area under the precision recall curve; DR, Diabetic retinopathy; NPDR, Non-proliferative diabetic retinopathy; PDR, Proliferative diabetic retinopathy; AMD, Age-related macular degeneration; MMD, Myopic macular degeneration

**Table S8. Diagnostic performance of Meta-EyeFM in detecting common ocular signs in internal testing**

|  | AUC, % (95% CI) | AUPRC (%) | Accuracy (%) | Sensitivity (%) | Specificity (%) | F1 score |
|---|---|---|---|---|---|---|
| **DR** | | | | | | |
| • Microaneurysm | 82.0 (78.6, 85.3) | 77.2 | 76.3 | 74.5 | 77.2 | 0.683 |
| • Laser scar | 99.0 (98.1, 100) | 92.5 | 96.4 | 95.7 | 96.4 | 0.759 |
| • Hard exudates | 95.6 (93.3, 98.0) | 83.6 | 90.8 | 88.3 | 91.0 | 0.883 |
| • Soft exudates | 93.4 (89.7, 97.1) | 59.3 | 81.9 | 90.9 | 81.5 | 0.302 |
| • Macular oedema | 98.7 (97.6, 99.7) | 86.2 | 93.9 | 94.1 | 93.9 | 0.577 |
| **AMD** | | | | | | |
| • Drusen | 77.9 (74.4, 81.4) | 96.5 | 83.0 | 85.8 | 56.3 | 0.901 |
| • Pigmentary abnormalities | 89.3 (86.6, 91.4) | 57.1 | 85.3 | 75.5 | 86.3 | 0.483 |
| **MMD** | | | | | | |
| • Tessellation | 95.3 (94.2, 96.5) | 97.6 | 87.5 | 86.9 | 88.8 | 0.905 |
| • Diffuse retinal atrophy | 97.1 (96.2, 97.9) | 76.7 | 90.9 | 96.2 | 90.3 | 0.679 |

Acronym: AUC, Area under the receiver operating curve; CI, Confidence interval; AUPRC, Area under the precision recall curve; DR, Diabetic retinopathy; AMD, Age-related macular degeneration; MMD, Myopic macular degeneration

**Table S9. Classification performance of Meta-EyeFM and different open-source LLM using different prompt methods to detect ocular diseases in external test dataset**

|  | Meta-EyeFM | Gemini-1.5 (1) | Gemini-1.5 (2) | Gemini-1.5 (3) | GPT-4o (1) | GPT-4o (2) | GPT-4o (3) |
|---|---|---|---|---|---|---|---|
| **CGMH (MMD)** | | | | | | | |
| Accuracy (%) | 91.5 | 50.5 | 69.5 | 80.5 | 50.0 | 55.0 | 64.0 |
| Sensitivity (%) | 91.0 | 10.0 | 95.0 | 78.0 | 0.0 | 13.0 | 98.0 |
| Specificity (%) | 92.0 | 100 | 44.0 | 83.0 | 100 | 97.0 | 30.0 |
| F1-Score | 0.915 | 0.020 | 0.757 | 0.800 | 0.000 | 0.224 | 0.731 |
| **IDRID (DR)** | | | | | | | |
| Accuracy (%) | 86.5 | 65.5 | 62.5 | 64.0 | 80.5 | 73.0 | 76.5 |
| Sensitivity (%) | 76.0 | 39.0 | 37.0 | 73.0 | 67.0 | 56.0 | 60.0 |
| Specificity (%) | 97.0 | 92.0 | 88.0 | 55.0 | 94.0 | 90.0 | 93.0 |
| F1-Score | 0.849 | 0.531 | 0.497 | 0.670 | 0.775 | 0.675 | 0.719 |
| **Pioneer (AMD)** | | | | | | | |
| Accuracy (%) | 76.0 | 49.5 | 64.0 | 58.0 | 51.0 | 57.5 | 62.0 |
| Sensitivity (%) | 78.0 | 2.0 | 71.0 | 79.0 | 3.0 | 24.0 | 77.0 |
| Specificity (%) | 73.0 | 97.0 | 57.0 | 37.0 | 99.0 | 91.0 | 47.0 |
| F1-Score | 0.767 | 0.038 | 0.664 | 0.653 | 0.058 | 0.361 | 0.670 |
| **Pioneer (Glaucoma)** | | | | | | | |
| Accuracy (%) | 80.5 | 48.5 | 62.0 | 58.0 | 50.5 | 53.0 | 57.5 |
| Sensitivity (%) | 66.0 | 2.0 | 52.0 | 41.0 | 6.0 | 19.0 | 89.0 |
| Specificity (%) | 95.0 | 95.0 | 72.0 | 75.0 | 95.0 | 87.0 | 26.0 |
| F1-Score | 0.838 | 0.037 | 0.578 | 0.494 | 0.108 | 0.288 | 0.677 |

Acronym: CGMH, Chang Gung Memorial Hospital; IDRiD, Indian Diabetic Retinopathy Dataset; PIONEER, Population Health and Age-related Sensory Decline Profile; MMD, Myopic macular degeneration; DR, Diabetic retinopathy; AMD, Age-related macular degeneration; Acc, Accuracy; Se, Sensitivity; Sp, Specificity.

Footnote: (1) denotes prompt "what is the disease in this image"; (2) denotes prompt "given a label set (DR, AMD, MMD, glaucoma): what is the disease in this image"' (3) denotes prompt that asked for the presence of a specific disease (e.g., Is DR or AMD or MMD or glaucoma presents in this image".

Table S10. Diagnostic performance of Meta-EyeFM and human graders in detecting ocular diseases in the internal test dataset

|  | Meta-EyeFM | Ophthal 1 | Ophthal 2 | Ophthal 3 | Ophthal 4 | Optom 1 | Optom 2 | Optom 3 | Optom 4 |
|---|---|---|---|---|---|---|---|---|---|
| **Normal (n= 16)** | | | | | | | | | |
| Accuracy (%) | 97.1 | 91.2 | 91.2 | 91.2 | 92.6 | 89.7 | 89.7 | 88.2 | 89.7 |
| Sensitivity (%) | 93.8 | 75.0 | 75.0 | 75.0 | 81.3 | 68.8 | 68.8 | 56.3 | 62.5 |
| Specificity (%) | 98.1 | 96.2 | 96.2 | 96.2 | 96.2 | 96.2 | 96.2 | 98.1 | 98.1 |
| F1-Score | 0.938 | 0.800 | 0.800 | 0.800 | 0.839 | 0.759 | 0.759 | 0.692 | 0.741 |
| **DR (n= 9)** | | | | | | | | | |
| Accuracy (%) | 98.5 | 92.6 | 98.5 | 97.1 | 94.1 | 95.6 | 100 | 91.2 | 97.1 |
| Sensitivity (%) | 88.9 | 88.9 | 88.9 | 88.9 | 66.7 | 88.9 | 100 | 88.9 | 88.9 |
| Specificity (%) | 100 | 93.2 | 100 | 98.3 | 98.3 | 96.6 | 100 | 91.5 | 98.3 |
| F1-Score | 0.941 | 0.762 | 0.941 | 0.889 | 0.750 | 0.842 | 1.000 | 0.727 | 0.889 |
| **AMD (n= 12)** | | | | | | | | | |
| Accuracy (%) | 91.2 | 92.6 | 92.6 | 94.1 | 95.6 | 91.2 | 91.2 | 94.1 | 89.7 |
| Sensitivity (%) | 83.3 | 75.0 | 66.7 | 100 | 100 | 75.0 | 100 | 66.7 | 83.3 |
| Specificity (%) | 92.9 | 96.4 | 100 | 92.9 | 94.6 | 94.6 | 89.3 | 100 | 91.1 |
| F1-Score | 0.769 | 0.783 | 0.800 | 0.857 | 0.889 | 0.750 | 0.800 | 0.800 | 0.741 |
| **MMD (n= 12)** | | | | | | | | | |
| Accuracy (%) | 98.5 | 97.1 | 91.2 | 100 | 97.1 | 97.1 | 97.1 | 98.5 | 95.6 |
| Sensitivity (%) | 91.7 | 91.7 | 91.7 | 100 | 83.3 | 100 | 91.7 | 100 | 100 |
| Specificity (%) | 100 | 98.2 | 91.1 | 100 | 100 | 96.4 | 98.2 | 98.2 | 94.6 |
| F1-Score | 0.957 | 0.917 | 0.786 | 1.000 | 0.909 | 0.923 | 0.917 | 0.960 | 0.889 |
| **Cataract (n= 9)** | | | | | | | | | |
| Accuracy (%) | 94.1 | 98.5 | 100 | 100 | 98.5 | 98.5 | 100 | 98.5 | 100 |
| Sensitivity (%) | 100 | 88.9 | 100 | 100 | 100 | 100 | 100 | 100 | 100 |
| Specificity (%) | 93.2 | 100 | 100 | 100 | 98.3 | 98.3 | 100 | 98.3 | 100 |
| F1-Score | 0.818 | 0.941 | 1.000 | 1.000 | 0.947 | 0.947 | 1.000 | 0.947 | 1.000 |
| **Glaucoma (n= 10)** | | | | | | | | | |
| Accuracy (%) | 89.7 | 88.2 | 85.3 | 89.7 | 82.4 | 82.4 | 85.3 | 83.8 | 82.4 |
| Sensitivity (%) | 80.0 | 20.0 | 0 | 40.0 | 20.0 | 10.0 | 0 | 10.0 | 0 |
| Specificity (%) | 91.4 | 100 | 100 | 98.3 | 93.1 | 94.8 | 100 | 96.6 | 96.6 |
| F1-Score | 0.696 | 0.333 | 0 | 0.706 | 0.250 | 0.143 | 0 | 0.154 | 0 |

Acronym: Ophthal, Ophthalmologist; Optom, Optometrist

**Annex A: Meta-EyeFM model development**

**Large language model training**

Figure A: Overview of large language model in Meta-EyeFM

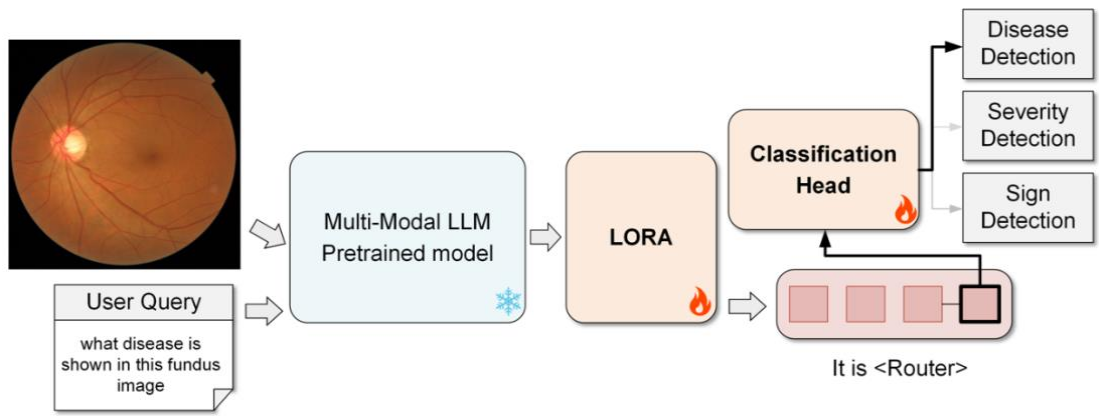

Routing Mechanism: To enhance the LLM with routing capabilities, we introduced the embedding-as-router paradigm. This involves expanding the original language model's vocabulary with a novel token <Router> which serves as a marker for the routing prediction task. Upon receiving a user text query $X_t$ and a corresponding fundus image $X_{img}$, the multimodal LLM F processes these inputs to produce a text response $\hat{y}$ expressed as $\hat{y} = F(X_{img}, X_t)$. When a routing decision is required, the output $\hat{y}$ includes the <Router> token. The LLM's final-layer embedding h_router, associated with the <Router> token is extracted and processed through a multi-layer perceptron (MLP) projection layer to yield the routing results $\hat{y}_{router} \in R^D$. Here $D$ denotes the number of expert models. The $i^{th}$ routing probability $\hat{y}_{router}^i$ is optimized using Binary Cross Entropy (BCE) loss. Our Large Language Model (LLM) is trained based on LLaVA1, using a parameter-efficient approach with Low-Rank Adaptation (LoRA). The training of the LLM is conducted end-to-end, combining text generation loss $L_T$ with Router's Binary Cross Entropy loss $L_R$. The total loss is computed as a weighted sum of these components, with weights $\lambda_T$ and $\lambda_R$: $L = \lambda_T L_T + \lambda_R L_R$.

**Vision foundation model training**

The encoder accepts unmasked image patches of size 16 × 16 pixels and projects them into feature vectors with a dimension of 1024. These feature vectors are processed through 24 Transformer blocks, each incorporating multiheaded self-attention and multilayer perceptron layers to produce high-level feature representations. The decoder then reintroduces masked dummy patches into these high-level features, subsequently reconstructing the original image patches through a linear projection. The training objective for the model was to reconstruct retinal images from their highly masked versions, employing a mask ratio of 0.75 for the fundus images.